\pdfoutput=1
\documentclass[a4paper,UKenglish,cleveref, autoref, thm-restate]{lipics-v2021}

\hideLIPIcs  

\title{PACE Solver Description: A Heuristic Directed Feedback Vertex Set Problem Algorithm} 

\author{Andrei Arhire}{Alexandru Ioan Cuza University of Iași, Romania\and }{andrei.arhire@info.uaic.ro}{}{}

\author{Paul Diac}{Alexandru Ioan Cuza University of Iași, Romania}{paul.diac@info.uaic.ro}{}{}

\authorrunning{A. Arhire, P. Diac} 

\Copyright{Andrei Arhire and Paul Diac} 



\category{} 

\relatedversion{} 

\supplement{The source code is available on Zenodo \href{https://zenodo.org/record/6646187}{(10.5281/zenodo.6646187)} and GitHub \href{https://github.com/AndreiiArhire/PACE2022}{ (https://github.com/AndreiiArhire/PACE2022)}.}

\nolinenumbers 

\EventEditors{John Q. Open and Joan R. Access}
\EventNoEds{2}
\EventLongTitle{42nd Conference on Very Important Topics (CVIT 2016)}
\EventShortTitle{CVIT 2016}
\EventAcronym{CVIT}
\EventYear{2016}
\EventDate{December 24--27, 2016}
\EventLocation{Little Whinging, United Kingdom}
\EventLogo{}
\SeriesVolume{42}
\ArticleNo{23}

\begin{document}

\maketitle

\begin{abstract}
A feedback vertex set of a graph is a set of nodes with the property that every cycle contains at least one vertex from the set i.e. the removal of all vertices from a feedback vertex set leads to an acyclic graph. In this short paper, we describe the algorithm for finding a minimum directed feedback vertex set used by the \texttt{\_UAIC\_ANDREIARHIRE\_} solver, submitted to the heuristic track of the 2022 PACE challenge. 
\end{abstract}

\section{Preliminaries}
\label{sec:typesetting-summary}

Let $G = (V, E)$ be a directed graph.
\begin{itemize}
    \item $N^{-}_{G}(v)= \{u \in V \mid \exists (u, v) \in E \wedge  \nexists (v, u) \in E \}$, \hspace{20px} $d^{-}_{G}(v) = |N^{-}_{G}(v)|$;
    \item $N^{+}_{G}(v)= \{u \in V \mid \nexists (u, v) \in E \wedge  \exists (v, u) \in E \}$, \hspace{20px} $d^{+}_{G}(v) = |N^{+}_{G}(v)|$;
    \item $N^{\pm}_{G}(v)= \{u \in V \mid \exists (u, v) \in E \wedge  \exists (v, u) \in E \}$, \hspace{20px} $d^{\pm}_{G}(v) = |N^{\pm}_{G}(v)|$;
    \item $G - v = (V \backslash \{v\}, E \cap \{\{V \backslash \{v\}\} \times \{V \backslash \{v\}\}\})$;
    \item \( G \circ v \) = $(V \backslash \{v\}, E \cap \{\{\{V \backslash \{v\}\} \times \{V \backslash \{v\}\}\} \cup \{\{N^{-}_{G}(v) \cup	 N^{\pm}_{G}(v)\} \times  \{N^{+}_{G}(v) \cup N^{\pm}_{G}(v)\}\}\})$;
    \item $G$ is a diclique if $\forall x, y \in V, x \neq y$ and $\{(x, y), (y, x)\} \subset E$.
\end{itemize}
We will refer to $G - v$ operation as vertex $v$ removal (remove $v$ together with all its adjacent edges) and to $G \circ v$ operation as vertex $v$ merger (connect all its predecessors with all its successors and exclude $v$ from the graph).

\section{Solver Summary}

The algorithm used by our solver has three main stages. In the first stage, it finds an initial solution. The second stage consists of removing redundant vertices contained in the initial solution. In the third stage, several local searches are performed based on the best-known solution.   

\section{Reduction Rules}

The problem can be viewed in the following way. We have to place each vertex in a set A or a set B such that at the end, A is a minimum feedback vertex set and B is the acyclic remainder. A group of vertices can be placed in A or B based on some verification which can be done in a polynomial time. 

\newpage

The following two operations are considered to be reductions:
\begin{itemize}
\item If a vertex can be part of set A  without affecting the solution's optimality, we remove it and eventually introduce it in the solution set.
\item If a vertex can be part of set B without affecting the solution's optimality, then we merge it. 
\end{itemize}
We make usage of 8 reduction rules described in \cite{DBLP:journals/Reductions1}, \cite{DBLP:journals/Reductions2} and \cite{DBLP:journals/Reductions3}. 

\newtheorem{reduction}[theorem]{Reduction Rule}

\begin{reduction}\label{testenv-reduction}
If there exists a vertex $v \in V $ and an edge $(v, v) \in E$, remove $v$.
\end{reduction}
In this case, vertex $v$ contains a self-loop. It is erased and inserted into the solution.
\begin{reduction}
If there exists a vertex $v \in V $, $(v, v) \not \in E$ with $d_{G}^{-}(v) + d_{G}^{\pm}(v) \leq 1$ or $d_{G}^{+}(v) + d_{G}^{\pm}(v) \leq 1$, merge $v$.
\end{reduction}
Vertex $v$ can have $d_{G}^{+}(v) = 0$ or $d_{G}^{-}(v) = 0$, then it can not be part of any cycle. Otherwise, $d_{G}^{+}(v) = 1$ or $d_{G}^{-}(v) = 1$, connecting all its predecessors with all its successors and excluding $v$ from the graph (merge operation) will not affect the optimality. In both scenarios $v$ is not considered in the solution.
\begin{reduction}
If there exists a vertex $v \in V $, $(v, v) \not \in E$, $\min{( d_{G}^{-}(v), d_{G}^{+}(v))} = 0$ and $N_{G}^{\pm}(v)$ forms a diclique, remove $N^{\pm}_{G}(v)$ and merge $v$.
\end{reduction}
The critical observation here is that in a diclique, at most, one node will not be part of the solution because any two vertices form a cycle. Thus, the diclique can be viewed as a single node, so rule 2 can be applied further.
\begin{reduction}
If there exist vertices $u$,  $v \in V $, $(u, v) \in E \wedge (v, u) \not\in E$ and $u$ and $v$ are in different strongly connected components of the graph $(V, E \backslash  \{  E \cap \{ (x,y), \forall x \in V, y \in N^{\pm}_{G}(x) \} \})$, erase $(u, v)$.
\end{reduction}
Edges between vertices in different strongly connected components are not part of any cycle (otherwise, the vertices would be in the same component). Thus all these edges can be deleted. Furthermore, in a diclique with two vertices, at least one node will be removed, so edges that are part of a diclique can be ignored when strongly connected components are computed. 
\begin{reduction}
If there exist vertices $u$, $v \in V $, $(u, v) \in E \wedge (v, u) \not\in E$ and $(N^{-}_{G}(u) \subset \{N^{-}_{G}(v) \cup N^{\pm}_{G}(v)\} ) \vee  (N^{+}_{G}(v) \subset \{N^{+}_{G}(u) \cup N^{\pm}_{G}(u)\})$, erase $(u, v)$.
\end{reduction}

The idea with this rule is to delete a set of edges with the property that there is no minimal cycle using them since only minimal cycles need to be broken to compute the feedback vertex set. 

The following three reduction rules are obtained based on the reduction rule 3 together with the idea that, at most, one node in a diclique is not part of the solution. 

\begin{reduction}
If there exists a node $v \in V $, $(v, v) \not\in E$ such that $\{N^{+}_{G}(v) \cup N^{\pm}_{G}(v)\}$ or $\{N^{-}_{G}(v) \cup N^{\pm}_{G}(v)\}$ forms a diclique, merge $v$.
\end{reduction}
\begin{reduction}
If there exists a node $v \in V $, $(v, v) \not\in E$ and $\{ N^{-}_{G}(v) \cup N^{+}_{G}(v) \cup N^{\pm}_{G}(v) \}$ can be split in two sets, $N^{\pm}_{G}(v)$ is included in one of them and each set forms a diclique, merge $v$.
\end{reduction}
\begin{reduction}
If there exists a node $v \in V $, $(v, v) \not\in E$, $d^{\pm}_{G}(v) = 0$ and $\{N^{+}_{G}(v) \cup N^{-}_{G}(v)\}$  can be split in at most three sets and each set forms a diclique, merge $v$.
\end{reduction}

\section{Stage one}
In first part of the algorithm reductions rules are applied until the graph no more supports any. If graph is not empty a promising vertex is selected to be part of the solution and it is removed. These two operations are performed until the graph becomes empty. A vertex $v$ is considered the best candidate in the selection process if it maximizes among all the other vertices either $(d^{+}_{G}(v)+d^{\pm}_{G}(v)) \cdot (d^{-}_{G}(v) + d^{\pm}_{G}(v))$ or $d^{\pm}_{G}(v) \cdot {\infty} + d^{-}_{G}(v) \cdot d^{+}_{G}(v)$.

\section{Stage two}
After many vertex selections, some of them may become redundant in the solution, i.e., they are not part of any cycle so they can be excluded from the feedback vertex set. To maximize the number of excluded vertices, we take them in the reversed order of their insertion and introduce them in the acyclic graph. At a fixed vertex, we check if the resulting graph is acyclic or not using an optimized version of the Tarjan algorithm for computing strongly connected components. If the graph is acyclic, the vertex will be erased from the solution, and the graph will keep the changes. This step is presented in \cite{DBLP:journals/Reductions3} and \cite{DBLP:journals/stage2}.
 
\section{Stage three}
 
 In the last part of the algorithm, some local searches are performed \cite{DBLP:journals/Reductions3}. A local search consists of running the same process as the one presented in stage one but on a subgraph of G. Subgraphs are obtained by removing a specific subset uniformly random from the best feedback vertex set found so far together with a specific subset from the acyclic remainder. Additionally, a slightly modified version of the algorithm from stage two is applied to the new solution depending on the remaining time.

\section{Complexity Analysis}

Both selection criteria are simple but not always optimal, adding significant benefits. Implying only the number of adjacent neighbors allows some reductions to be performed efficiently using hash sets, keeping most of the data in memory. After each update of the graph, reductions $1$ and $2$ can be applied in $\mathcal{O}(1)$. Reduction $3$ is implemented in $\mathcal{O}((|V|+|E|) \cdot \log{(|V|)})$ and reduction $4$ in $\mathcal{O}(|V|+|E|)$. The rest are run only for vertices with degrees bounded by a small constant, and we assume the complexity is $\mathcal{O}(|V|+|E|)$.  
The first two reductions are run after each graph change. The others are run after each loss of around $5\%-25\%$ of the number of edges. For every iteration among $T$ ones, we choose to restore around 30\% of the vertices into the acyclic remainder and run the local search. As an instance can be reduced in linear time to a graph with $ |V| < |E| $, the final complexity of the program is $\mathcal{O}(T \cdot (|V|+|E| \cdot \log{|E|}))$, based on Master's Theorem \cite{Verma-1994}, where $T$ is the number of local searches. $T$ is a parameter bounded by the time limit.

\newpage

\bibliography{lipics-v2021-AFC}

\begin{thebibliography}{1}

\bibitem{DBLP:journals/stage2}
Berend Hasselman.
\newblock {An efficient method for detecting redundant feedback vertices}.
\newblock CPB Discussion Paper~29, CPB Netherlands Bureau for Economic Policy
  Analysis, April 2004.
\newblock URL: \url{https://ideas.repec.org/p/cpb/discus/29.html}.

\bibitem{DBLP:journals/Reductions2}
Hen-Ming Lin; Jing-Yang Jou.
\newblock On computing the minimum feedback vertex set of a directed graph by
  contraction operations.
\newblock {\em IEEE Transactions on Computer-Aided Design of Integrated
  Circuits and Systems}, 19:295--307, 2000.
\newblock URL: \url{http://doi.org/10.1109/43.833199}, \href
  {https://doi.org/10.1109/43.833199} {\path{doi:10.1109/43.833199}}.

\bibitem{DBLP:journals/Reductions3}
Mile Lemaic.
\newblock {\em Markov-Chain-Based Heuristics for the Feedback Vertex Set
  Problem for Digraphs}.
\newblock PhD thesis, Universit{\"a}t zu K{\"o}ln, 2008.

\bibitem{DBLP:journals/Reductions1}
Hanoch Levy; David~W Low.
\newblock A contraction algorithm for finding small cycle cutsets.
\newblock {\em Journal of Algorithms}, 9:470--493, 1988.
\newblock URL: \url{http://doi.org/10.1016/0196-6774%2888%2990013-2}, \href
  {https://doi.org/10.1016/0196-6774(88)90013-2}
  {\path{doi:10.1016/0196-6774(88)90013-2}}.

\bibitem{Verma-1994}
R.M. Verma.
\newblock A general method and a master theorem for divide-and-conquer
  recurrences with applications.
\newblock {\em Journal of Algorithms}, 16:67--79, 1994.
\newblock URL: \url{http://doi.org/10.1006/jagm.1994.1004}, \href
  {https://doi.org/10.1006/jagm.1994.1004} {\path{doi:10.1006/jagm.1994.1004}}.

\end{thebibliography}

\end{document}